# Fast Super-resolution 3D SAR Imaging Using an Unfolded Deep Network

Jingkun Gao, *Student member, IEEE*, Bin Deng, Yuliang Qin, Hongqiang Wang and Xiang Li

*Abstract*— For 3D Synthetic Aperture Radar (SAR) imaging, one typical approach is to achieve the cross-track 1D focusing for each range-azimuth pixel after obtaining a stack of 2D complex-valued images. The cross-track focusing is the main difficulty as its aperture length is limited and the antenna positions are usually non-uniformly distributed. Sparsity regularization methods are widely used to tackle these problems. However, these methods are of obvious limitations. The most well-known ones are their heavy computational burdens and unsatisfied stabilities. In this letter, an efficient deep network-based cross-track imaging method is proposed. When trained, the imaging process, i.e. the forward propagation of the network, is made up of simple matrix-vector calculations and element-wise nonlinearity operations, which significantly speed up the imaging. Also, we find that the deep network is of good robustness against noise and model errors. Comprehensive simulations and experiments have been carried out, and the superiority of the proposed method can be clearly seen.

*Index Terms*— **3D Synthetic Aperture Radar (3D SAR), line spectral estimation, deep networks, tomography SAR (TomoSAR), down-looking linear array SAR (DLLA-SAR).**

## I. INTRODUCTION

THREE dimension SAR overcomes some drawbacks of conventional SAR and can obtain more essential information of the target. Typical 3D SAR regimes include TomoSAR [1], DLLA-SAR [2] and Circular SAR (CSAR) [3]. Although their observation geometries differ a lot from each other, they share one common imaging approach, i.e. 2D imaging followed by 1D focusing of the third. There exists more general 3D SAR imaging approaches [4],[5], while we mainly focus on those employ the "2D+1D" approach in this letter. As the cross-track aperture size is usually much shorter than that of the along-track, the resolution is not satisfied and super-resolution focusing is needed. With the development of sparse recovery algorithms and the Compressive Sensing (CS) [6] theory, great progress has been made on the super-resolution third dimension focusing in 3D SAR.

Zhu et al. [7] carried out series of studies on TomoSAR imaging of urban areas. They introduced the CS approach to the elevation focusing, and impressive 3D maps of urban areas were presented. Also, the super-resolution power of CS-based methods for TomoSAR was analyzed in depth [8]. Reigber et al. [9] investigated the focusing problem towards volume scattering objects such as forests. They proposed an imaging method that combined the wavelet representation and

Manuscript received …., 2018. Project supported by the National Natural Science Foundation of China (Grant No. 61571011 and 61701513).

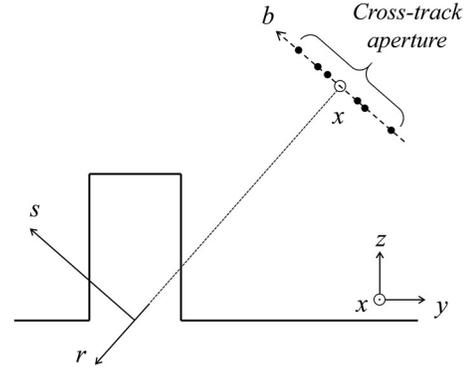

Fig. 1. Simplified TomoSAR geometry.

sparse-recovery. Peng et al. [10] utilized L1-regularization to achieve cross-track super-resolving for DLLA-SAR. Bao et al. [11] suggested a gridless sparse recovery method to tackle the off-grid mismatch problem. Zhang et al. [12] recently proposed a method based on matrix completion for DLLA-SAR. For CSAR, Austin et al. [3] applied sparse signal methods to trajectories with different elevations to form 3D images. As can be seen, sparsity-driven algorithms have been widely used in 3D SAR. However, a well-known drawback of these algorithms is their heavy computational burden. Usually, a large number of iterations are needed, and some time-consuming operations such as matrix inversion may exist in each iteration that will further extend the time costs. Secondly, the stability of these algorithms is sometimes troublesome. As pointed out by [8], noise or model errors can easily lead to spurious artifacts.

The cross-track focusing is essentially a line spectral estimation problem which belongs to a kind of nonlinear regression problem. Deep networks can just act as a general and powerful nonlinear mapping function [13]. In this letter, we propose a deep network-based line spectral estimation method and introduce it to 3D SAR. The superiority of the proposed algorithm is verified by both simulations and experiments.

## II. BACKGROUND

### A. Signal Model

For the convenience, we will take the simplified TomoSAR geometry as an example (Fig. 1). Suppose the range-azimuth 2D imaging, image registration and deramping have been finished, then the cross-track signal for a range-azimuth pixel at $(x, r)$ can be expressed as

$$g(b) = \int \exp\left(2j\frac{k_c}{r}bs\right) \cdot \gamma(s)ds \qquad (1)$$

## Algorithm 1: The VAMP algorithm.

Initialize $\alpha_{2,0} = \left\langle \boldsymbol{\eta}_2'\left(\boldsymbol{v}_{2,0}, \boldsymbol{\chi}_{2,0}, \boldsymbol{\theta}_0\right)\right\rangle \boldsymbol{\theta}_0$, $\boldsymbol{\chi}_{2,0} = \boldsymbol{\beta}_0 \|\boldsymbol{g}\|_2^2 / N$,
$\boldsymbol{v}_{2,0} = \boldsymbol{R}_0 \hat{\boldsymbol{g}}$, , $\hat{\boldsymbol{\gamma}}_0 = \boldsymbol{\eta}_2\left(\boldsymbol{v}_{2,0}, \boldsymbol{\chi}_{2,0}, \boldsymbol{\theta}_0\right)$, , where $\boldsymbol{\chi}_{2,0}, \boldsymbol{\beta}_0, \boldsymbol{v}_{2,0} \in \mathbb{R}^{2M}$,
$\boldsymbol{R}_0 \in \mathbb{R}^{2M \times 2N}$, $\boldsymbol{\theta}_0 \in \mathbb{R}^{2M \times D}$ and $D$ is determined by $\boldsymbol{\eta}_2(\bullet)$.

For t = 1,2,…,T

  1: $\boldsymbol{v}_{1,t} = \left(\hat{\boldsymbol{\gamma}}_{t-1} - \alpha_{2,t-1} \boldsymbol{v}_{2,t-1}\right) / \left(1 - \alpha_{2,t-1}\right)$

  2: $\boldsymbol{\chi}_{1,t} = \alpha_{2,t-1} \boldsymbol{\chi}_{2,t-1} / \left(1 - \alpha_{2,t-1}\right)$

  3: $\hat{\boldsymbol{\gamma}}_t = \boldsymbol{\eta}_1\left(\boldsymbol{v}_{1,t}, \boldsymbol{G}_t, \boldsymbol{R}_t\right)$

  4: $\alpha_{1,t} = \left\langle \boldsymbol{\eta}_1'\left(\boldsymbol{v}_{1,t}, \boldsymbol{G}_t, \boldsymbol{R}_t\right)\right\rangle$

  5: $\boldsymbol{v}_{2,t} = \left(\hat{\boldsymbol{\gamma}}_t - \alpha_{1,t} \boldsymbol{v}_{1,t}\right) / \left(1 - \alpha_{1,t}\right)$

  6: $\boldsymbol{\chi}_{2,t} = \alpha_{1,t} / \left(1 - \alpha_{1,t}\right) \cdot \boldsymbol{\chi}_{1,t} \odot \boldsymbol{\beta}_t$

  7: $\hat{\boldsymbol{\gamma}}_t = \boldsymbol{\eta}_2\left(\boldsymbol{v}_{2,t}, \boldsymbol{\chi}_{2,t}, \boldsymbol{\theta}_t\right)$

  8: $\alpha_{2,t} = \left\langle \boldsymbol{\eta}_2'\left(\boldsymbol{v}_{2,t}, \boldsymbol{\chi}_{2,t}, \boldsymbol{\theta}_t\right)\right\rangle$

Return $\hat{\boldsymbol{\gamma}}_T$

▨ Algorithm steps    ☐ Trainable parameters

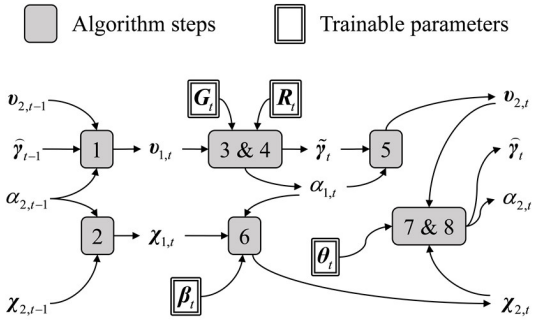

Fig. 2. One layer of the unfolded deep network

where $k_c = 2\pi / \lambda_c = 2\pi f_c / c$ is the center spatial wavenumber. The expected cross-track resolution for nonparametric spectral analysis is approximately [7]

$$\rho_s = \frac{\lambda_c r}{2\Delta b} \qquad (2)$$

where $\Delta b$ is the aperture length along axis $b$. In addition, the following constraint should be satisfied [7] for TomoSAR

$$\Delta s \ll \frac{\rho_r r}{\Delta b} \qquad (3)$$

where $\rho_r$ is the range resolution and $\Delta s$ represents the extent of the objects under illumination.

### B. Related Works

Discretize (1) and neglect the integral constant term

$$g_n = \sum_{m=1}^{M} \exp\left(2j \frac{k_c}{r} b_n s_m\right) \cdot \gamma_m, \quad n = 1, \dots, N \qquad (4)$$

where $N, M$ represent the number of cross-track acquisitions and the number of discrete $s$ samples respectively. Take noise into consideration and the cross-track imaging model becomes

$$\boldsymbol{g} = \boldsymbol{H}\boldsymbol{\gamma} + \boldsymbol{\omega}, \quad \boldsymbol{g}, \boldsymbol{\omega} \in \mathbb{C}^N, \boldsymbol{\gamma} \in \mathbb{C}^M, \boldsymbol{H} \in \mathbb{C}^{N \times M} \qquad (5)$$

where $\boldsymbol{\omega}$ is the noise, $\boldsymbol{H}$, with $H_{nm} = \exp\left(2jk_c b_n s_m / r\right)$, is the sensing matrix. Recently, the most effective approaches to solve $\boldsymbol{\gamma}$ are CS-based sparse recovery methods. Typically, the

following optimization problem is solved to reconstruct $\hat{\boldsymbol{\gamma}}$

$$\hat{\boldsymbol{\gamma}} = \arg\min_{\boldsymbol{\gamma}} \|\boldsymbol{g} - \boldsymbol{H}\boldsymbol{\gamma}\|_2^2 + \varepsilon \|\boldsymbol{\Gamma}\boldsymbol{\gamma}\|_1 \qquad (6)$$

where $\varepsilon$ is the Lagrange multiplier. We suppose $\boldsymbol{\gamma}$ is of sparsity directly and $\boldsymbol{\Gamma}$ is thought to be a unit matrix. Most of the algorithms for solving (6) consist of iterative-based processes which essentially limit the efficiency. We will see in the following sections that the proposed deep network-based method breaks through this bottleneck.

## III. METHOD

### A. The unfolded deep network

Recently, a method coined "deep unfolding" has been proposed to design deep network structures [14]. This method provides a good combination of hand designed algorithms and the ability of networks to learn from training data. We firstly give a brief introduction to the adopted sparse recovery algorithm to be unfolded, i.e. the Vector Approximate Massage Passing (VAMP) algorithm [15] (Algorithm 1). As it is originally designed for real-valued problems, we carry out the following adaptations to (5)

$$\hat{\boldsymbol{g}} = \hat{\boldsymbol{H}}\hat{\boldsymbol{\gamma}} + \hat{\boldsymbol{\omega}}, \quad \hat{\boldsymbol{g}}, \hat{\boldsymbol{\omega}} \in \mathbb{R}^{2N}, \hat{\boldsymbol{\gamma}} \in \mathbb{C}^{2M}, \hat{\boldsymbol{H}} \in \mathbb{C}^{2N \times 2M} \qquad (7)$$

where

$$\hat{\boldsymbol{H}} = \begin{bmatrix} \Re(\boldsymbol{H}) & -\Im(\boldsymbol{H}) \\ \Im(\boldsymbol{H}) & \Re(\boldsymbol{H}) \end{bmatrix}, \ \hat{\boldsymbol{g}} = \begin{bmatrix} \Re(\boldsymbol{g}) \\ \Im(\boldsymbol{g}) \end{bmatrix}, \ \hat{\boldsymbol{\gamma}} = \begin{bmatrix} \Re(\boldsymbol{\gamma}) \\ \Im(\boldsymbol{\gamma}) \end{bmatrix} \qquad (8)$$

and $\Re(\bullet), \Im(\bullet)$ represent the real and imaginary operators.

In Algorithm 1,

$$\boldsymbol{\eta}_1\left(\boldsymbol{v}_{1,t}, \boldsymbol{G}_t, \boldsymbol{R}_t\right) = \boldsymbol{G}_t \boldsymbol{v}_{1,t} + \boldsymbol{R}_t \hat{\boldsymbol{g}}, \quad \boldsymbol{G}_t \in \mathbb{R}^{2M \times 2M} \qquad (9)$$

and $\boldsymbol{\eta}_1'\left(\boldsymbol{v}_{1,t}, \boldsymbol{G}_t, \boldsymbol{R}_t\right)$ is defined as the diagonal of the Jacobian

$$\boldsymbol{\eta}_1'\left(\boldsymbol{v}_{1,t}, \boldsymbol{G}_t, \boldsymbol{R}_t\right) = \text{diag}\left[\frac{\partial \boldsymbol{\eta}_1\left(\boldsymbol{v}_{1,t}, \boldsymbol{G}_t, \boldsymbol{R}_t\right)}{\partial \boldsymbol{v}_{1,t}}\right] \qquad (10)$$

Given (10), we can know that $\left\langle \boldsymbol{\eta}_1'\left(\boldsymbol{v}_{1,t}, \boldsymbol{G}_t, \boldsymbol{R}_t\right)\right\rangle = \text{tr}\left(\boldsymbol{G}_t\right) / 2M$.

$\boldsymbol{\eta}_2(\bullet)$ represents an element-wise nonlinear transform which is physically interpreted as a denoiser. Here $\boldsymbol{\eta}_2(\bullet)$ is chosen to be the piecewise-linearity function defined in [15]. More details can be found in [15], and we will not repeat them here.

With the above definitions, we can unfold the algorithm to get the corresponding deep network [15]. Fig. 2 shows one layer of the unfolded network which consists $T$ layers in all. Actually, to make parameters $\boldsymbol{G}_t, \boldsymbol{R}_t, \boldsymbol{\beta}_t, \boldsymbol{\theta}_t$ trainable is one essential difference between the unfolded deep network and the original hand designed algorithm. In the original VAMP algorithm, these parameters are fixed or definitive. Making them trainable enables the network to get competitive or even better performance with less layers. After training, the feed-forward process consists only simple low-complexity calculations, which make the network more efficient.

### B. Train the network

To train the network, we need to generate enough training data $\left\{\left(\hat{\boldsymbol{g}}^{(p)}, \hat{\boldsymbol{\gamma}}^{(p)}\right)\right\}_{p=1}^{P}$, where $P$ represents the size of the training dataset and $\hat{\boldsymbol{g}}$, $\hat{\boldsymbol{\gamma}}$ are the real counterparts of $\boldsymbol{g}$, $\boldsymbol{\gamma}$ defined in (8). To imitate the real scattering process as fidelity as possible, the echo $\boldsymbol{g}$ is generated by (1) with continuously sampled $s$. For generating the corresponding ground truth image $\boldsymbol{\gamma}$, the continuous $s$ is approximated by the nearest discrete $s_m$. The number of point scatters is made to be uniformly distributed within $\{1,2,3,4\}$. For each scatter, its position $s$ is uniformly sampled within the extent $\Delta s$, and its scattering coefficient $\gamma$ is a complex Gaussian random number that satisfy $\gamma \sim N(0,1)+jN(0,1)$. The loss function of the network is defined as

$$\ell_T(\Theta) = \frac{1}{Q}\sum_{q=1}^{Q}\left\|\hat{\boldsymbol{\gamma}}_T\left(\hat{\boldsymbol{g}}^{(q)},\Theta\right)-\hat{\boldsymbol{\gamma}}^{(q)}\right\|_2^2 \qquad (11)$$

where $\Theta=\{\boldsymbol{G}_t, \boldsymbol{R}_t, \boldsymbol{\beta}_t, \boldsymbol{\theta}_t\}_{t=1}^{T}$ denotes all the trainable parameters, $\hat{\boldsymbol{\gamma}}_T\left(\hat{\boldsymbol{g}}^{(q)},\Theta\right)$ is the output of layer $T$ of the $q$-th input $\hat{\boldsymbol{g}}$, and $Q$ is the batch size. All the networks in this letter are trained with $P=200,000$, $Q=500$ and $T=8$. To avoid bad local minimum, we employ the layer-by-layer training strategy as suggested in [15]. The parameters are initialized as

$$\boldsymbol{\beta}_t = \mathbf{1}, \boldsymbol{R}_t = \bar{\boldsymbol{H}}^{\mathrm{T}}, \boldsymbol{G}_t = \boldsymbol{E} \qquad (12)$$

where $\bar{\boldsymbol{H}}$ is defined in (8) and $\boldsymbol{E}$ denotes the unit matrix. We implemented the network using Tensorflow, and a TITAN xp GPU was used to accelerate the training.

TABLE I
SIMULATION PARAMETERS

| Carrier frequency $f_c$ | 10 GHz | Bandwidth | 200 MHz |
|---|---|---|---|
| Frequency points | 801 | Range from radar to target | 800 km |
| Along-track beam width | 1° | Along-track sampling interval | 1 m |
| Cross-track aperture length $\Delta b$ | 300 m | Cross-track acquisitions $N$ | 31 |
| Cross-track image extent $\Delta s$ | 300 m | Cross-track pixels $M$ | 78 |

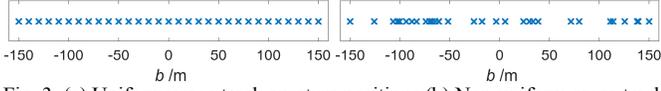

Fig. 3. (a) Uniform cross-track aperture positions (b) Non-uniform cross-track aperture positions.

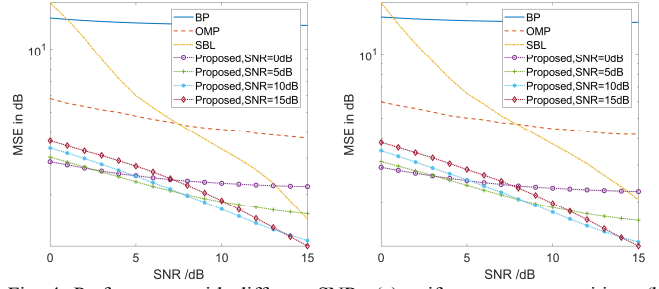

Fig. 4. Performance with different SNRs (a) uniform aperture positions (b) non-uniform aperture positions.

TABLE II
TIME NEEDS FOR 1000 TESTS OF DIFFERENT METHODS

| Algorithms | Time needs for 1000 tests |
|---|---|
| BP | 0.015 s |
| OMP | 0.496 s |
| SBL | 935.44 s |
| The proposed method | 0.191 s (GPU: 0.008 s) |

## IV. RESULTS

### A. Performance analysis

Given TABLE I, one can easily verify the constraint in (3). Based on (2), we can induce that the cross-track resolution is approximately 40 m. As the pixel interval along cross-track dimension is no larger than 4 m ($\Delta s/M$), we expect a 10-times super-resolution ability. We use both uniform and non-uniform antenna positions as shown in Fig. 3 to test the focusing performances of different algorithms.

We choose the Mean Square Error (MSE) as the quantitative index to evaluate the reconstruction precision. Three typical reconstruction algorithms, the nonparametric Back Projection (BP) algorithm (i.e. $\hat{\boldsymbol{\gamma}} = \boldsymbol{H}^{\mathrm{H}}\boldsymbol{g}$), the Orthogonal Matching Pursuit (OMP) algorithm [16] and the Sparse Bayesian

Learning (SBL) algorithm [17], were chosen to compete with the proposed method. For a given antenna position distribution, we train four networks under four different signal-to-noise ratios (SNR), i.e. 0 dB, 5 dB, 10 dB, 15 dB, respectively. Consequently, we will totally get 8 networks. The performances of different algorithms and networks are drawn in Fig. 4. For each SNR, 1000 tests was performed to evaluate the MSE. We can see that the performance of the deep network-based algorithms were competitive no matter for which SNR they were trained. Although the SBL algorithm show good performance in high SNRs, its performance degraded dramatically as SNR went lower. Compared with the BP, OMP and SBL, the networks trained under 10 dB and 15 dB show better performance for both aperture distributions and all SNR levels. One can note that when SNR = 0 dB, 5 dB, 10 dB, 15dB, the network trained under corresponding SNR show the best performance among all algorithms and networks. In fact, the SNR level can be estimated before imaging [7], and we can choose the appropriate network accordingly. As a result, an actual performance better than any individual network can be expected.

Moreover, we recorded the time needs for 1000 tests in TABLE II. These data were the average results among all SNR levels. The test platform was a PC with one Intel i3-4130 CPU. As aforementioned, the proposed deep network-based method can be easily parallelized. Therefore, we also tested the time needs of the proposed method with GPU implementations. The superiority of our method on imaging efficiency can be clearly seen especially when implemented on a GPU.

### B. 3D scene simulation

The height of the "building" in Fig. 5 (a) is 84 meters. The scattering coefficients of all the point scatters on the "building" surface are of unit intensity and uniform random phase within $(0, 2\pi)$. The elevation angle is $45°$. Other parameters are listed in TABLE I. Fig. 5 (b) draws the 2D SAR imaging results of the object in Fig. 5 (a). We can clearly see the layover effects. For 3D SAR imaging, we suppose the cross-track aperture is the non-uniform one shown in Fig. 3 (b).

In Fig. 6 (a), the cross-track resolution of the BP algorithm width of the projected two facets in Fig. 6 (h) are thinner than those in Fig. 6 (d) and (f) which indicates a better reconstruction quality. In Fig. 7, we projected the scatters onto z-axis, i.e. neglecting their x and y coordinates, and we draw the normalized histograms of the reconstructed z coordinates. We can see two peaks that represent the ground and the building roof respectively. With infinite elevation resolution, the scatters at the ground and the roof will be ideally gathered and two $\delta$ functions are expected. Accordingly, we can judge the focusing quality by comparing the sharpness of the curves. We can see that the sharpness of the proposed method is no weaker than other algorithms. Moreover, TABLE III recorded the time needs for above 3D reconstructions. Again, we can see the advantage of the proposed algorithm clearly.

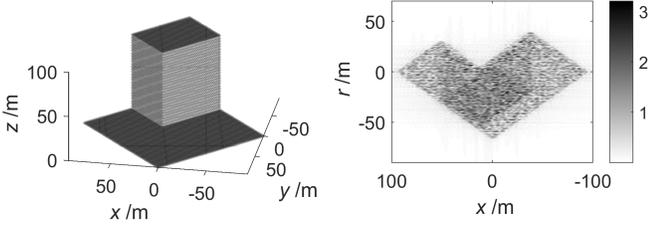

Fig. 5. (a) A "building"-shape object made up by hundreds of point scatters (b) conventional 2D SAR imaging results of the "building"-shape object.

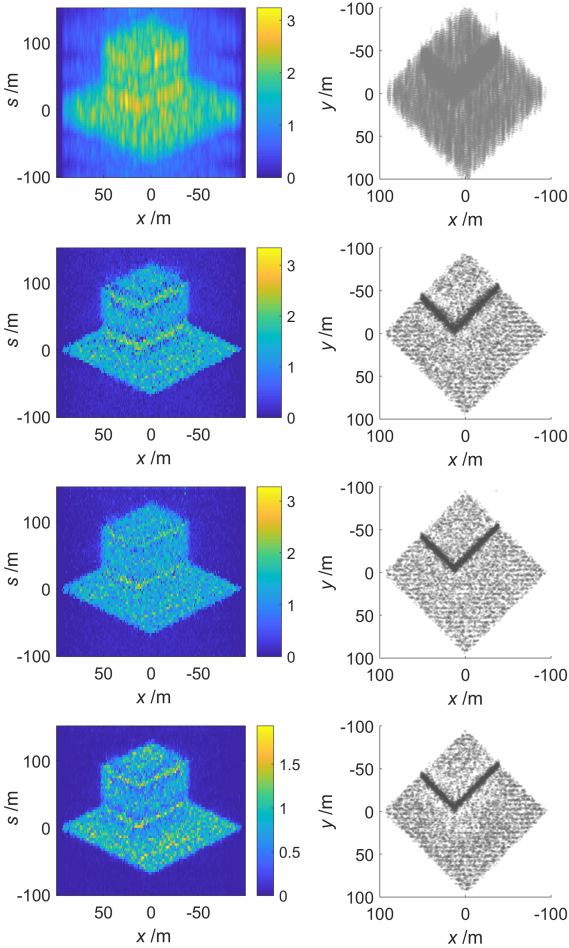

Fig. 6. 3D imaging results of different algorithms. Left column: Maximum projection of the 3D imaging results to the x-s plane; Right column: The projected "hot map" onto the x-y plane. Top to bottom: BP, OMP, SBL, the proposed method.

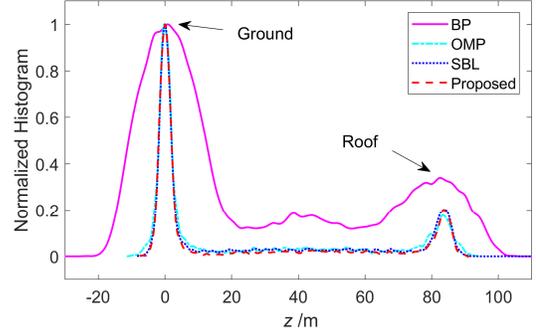

Fig. 7. Normalized histogram of the reconstructed z coordinates.

TABLE III
TIME NEEDS FOR THE 3D SCENE RECONSTRUCTION

| Algorithms | Time needs |
| --- | --- |
| BP | 0.18 s |
| OMP | 5.31 s |
| SBL | 9520.38 s |
| The proposed method | 2.26 s (GPU: 0.091 s) |

### C. Laboratory results

A proof-of-concept experiment was carried out in an electromagnetic anechoic chamber. The experimental scenarios are shown in Fig. 8, and the parameters are listed in TABLE IV. Similar with the 3D simulation results, the BP reconstruction is of low cross-track resolution while with high side-lobes. The other three algorithms all achieve super-resolved images. However, some spurious artifacts exist in Fig. 9 (b), and the image in Fig. 9 (c) is not as clear as the one in Fig. 9 (d). In addition, the proposed algorithm took about 0.012 s (GPU: 8 ms) to finish the focusing, while the OMP algorithm took 0.04s and the SBL took about 8 s.

### D. Discussion

Actually, we had also tried a widely used structure, i.e. convolutional neural network (CNN) [18], for the line spectral estimation problem in this letter. However, we got no satisfied results. In our opinion, CNN is powerful and suitable for *local transforms* as the receptive field is limited by the convolutional kernel. For the line spectral estimation problem, *global information* of the input signal is preferred to provide more accurate estimation. As a result, the deep architecture we employed here is a more reasonable and effective choice for 3D SAR application. Besides, there are some limitations in our recent method. To train the network, we now need the aperture was much lower than that of the along-track dimension. On the contrary, OMP, SBL and the proposed algorithm all provided super-resoled results. However, the background of Fig. 6 (g) was much clearer than those in Fig. 6 (c) and (e), and the magnitude of the image in Fig. 6 (g) revealed the scattering intensity more authentically. Also, Fig. 6 (g) has fewer outliers. To evaluate the quality of the 3D reconstruction results more intuitively, we projected the reconstructed 3D scatters to different axes or planes to show their statistical properties. On the right column of Fig. 6, we projected the reconstructed scatters onto the x-y plane, and the intensity of the "hot map" reveals the gathering extent of the scatters. Unsurprisingly, the "hot map" in Fig. 6 (b) is seriously blurred. In Fig. 6 (d), (f) and (h), we can see the reconstructed two facets of the building clearly. However, the positions are known in advance. Whether we can train one network to work with different antenna topologies and SNR levels is a meaningful topic for further research.

| Carrier frequency $f_c$ | 35 GHz | Bandwidth | 6 GHz |
|---|---|---|---|
| Frequency points | 501 | Range from radar to target | 3.3 m |
| Along-track beam width | 14° | Along-track sampling interval | 4 mm |
| Cross-track aperture size $\Delta b$ | 6 cm | Cross-track acquisitions $N$ | 31 |
| Cross-track image size $\Delta s$ | 60 cm | Cross-track pixels $M$ | 35 |

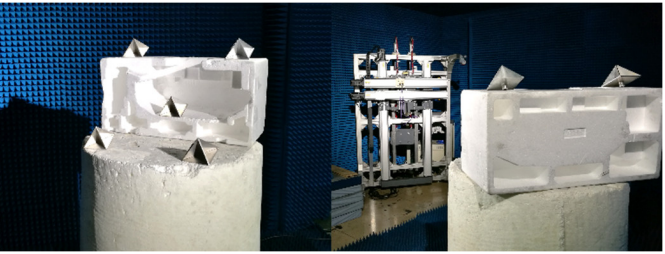

Fig. 8. Experimental scenarios.

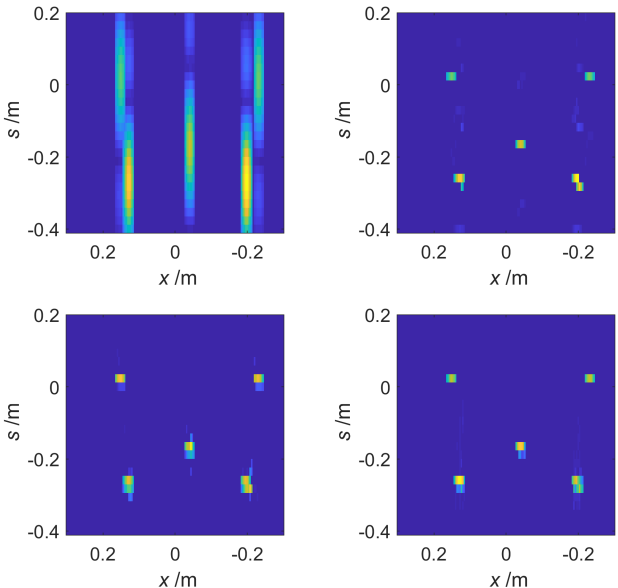

Fig. 9. Maximum projection of the 3D imaging results to the x-s plane, (a) BP (b) OMP (c) SBL (d) the proposed method.

## V. CONCLUSION

An efficient deep network-based line spectral estimation algorithm was proposed and applied to 3D SAR imaging. The structure design and training method of the deep network were firstly introduced. Compared with conventional iterative-based sparse regularization algorithms, the feed-forward of the network consists only simple calculations and can be easily parallelized. These facts greatly speed up the imaging process. Also, the learning capability of the network helps it achieve competitive or even better reconstruction performance. Simulations and experiments verified the superiority of the proposed method clearly. How to train one network to tackle different aperture topologies and SNR levels is an appealing future topic. Finally, we suggest that the proposed method can also be used for many line spectral estimation applications such as direct of arrival estimation.

## REFERENCES

[1] G. Fornaro, F. Lombardini and F. Serafino, "Three-dimensional multipass SAR focusing: experiments with long-term spaceborne data," *IEEE Transactions on Geoscience & Remote Sensing*, vol.43, no.4, pp. 702-714, 2005.

[2] J. Klare, M. Weiss, O. Peters and A. Brenner, "ARTINO: A New High Resolution 3D Imaging Radar System on an Autonomous Airborne Platform,"2006, pp. 3842-3845.

[3] C. D. Austin, E. Ertin and R. L. Moses, "Sparse Signal Methods for 3-D Radar Imaging," *IEEE Journal of Selected Topics in Signal Processing*, vol.5, no.3, pp. 408-423, 2011.

[4] J. Shi, X. Zhang, X. Gao and J. Jianyu, "Signal Processing for Microwave Array Imaging: TDC and Sparse Recovery," *IEEE Transactions on Geoscience & Remote Sensing*, vol.50, no.11, pp. 4584-4598, 2012.

[5] J. Gao, Y. Qin, B. Deng, H. Wang and X. Li, "Novel Efficient 3D Short-Range Imaging Algorithms for a Scanning 1D-MIMO Array," *IEEE Transactions on Image Processing*, vol.27, no.7, pp. 3631-3643, 2018.

[6] E. J. Candes, J. Romberg and T. Tao, "Robust uncertainty principles: exact signal reconstruction from highly incomplete frequency information," *IEEE Transactions on Information Theory*, vol.52, no.2, pp. 489-509, 2004.

[7] X. X. Zhu and R. Bamler, "Very High Resolution Spaceborne SAR Tomography in Urban Environment," *IEEE Transactions on Geoscience & Remote Sensing*, vol.48, no.12, pp. 4296-4308, 2010.

[8] X. X. Zhu and R. Bamler, "Super-Resolution Power and Robustness of Compressive Sensing for Spectral Estimation With Application to Spaceborne Tomographic SAR," *IEEE Transactions on Geoscience & Remote Sensing*, vol.50, no.1, pp. 247-258, 2011.

[9] E. Aguilera, M. Nannini and A. Reigber, "Wavelet-Based Compressed Sensing for SAR Tomography of Forested Areas," *IEEE Transactions on Geoscience & Remote Sensing*, vol.51, no.12, pp. 5283-5295, 2013.

[10] X. Peng, W. Tan, W. Hong, C. Jiang, Q. Bao and Y. Wang, "Airborne DLSLA 3-D SAR Image Reconstruction by Combination of Polar Formatting and $L\_1S$ Regularization," *IEEE Transactions on Geoscience & Remote Sensing*, vol.54, no.1, pp. 213-226, 2016.

[11] Q. Bao, X. Peng, Z. Wang, Y. Lin and W. Hong, "DLSLA 3-D SAR Imaging Based on Reweighted Gridless Sparse Recovery Method," *IEEE Geoscience & Remote Sensing Letters*, vol.13, no.6, pp. 841-845, 2017.

[12] S. Zhang, G. Dong and G. Kuang, "Matrix Completion for Downward-Looking 3-D SAR Imaging With a Random Sparse Linear Array," *IEEE Transactions on Geoscience & Remote Sensing*, vol.PP, no.99, pp. 1-13, 2017.

[13] K. H. Jin, M. T. Mccann, E. Froustey and M. Unser, "Deep Convolutional Neural Network for Inverse Problems in Imaging," *IEEE Transactions on Image Processing A Publication of the IEEE Signal Processing Society*, vol.26, no.9, pp. 4509-4522, 2017.

[14] J. R. Hershey, J. L. Roux and F. Weninger, "Deep Unfolding: Model-Based Inspiration of Novel Deep Architectures," *Computer Science*, 2014.

[15] M. Borgerding, P. Schniter and S. Rangan, "AMP-Inspired Deep Networks for Sparse Linear Inverse Problems," *IEEE Transactions on Signal Processing*, vol.65, no.16, pp. 4293-4308, 2016.

[16] J. A. Tropp and A. C. Gilbert, "Signal Recovery From Random Measurements Via Orthogonal Matching Pursuit," *IEEE Transactions on Information Theory*, vol.53, no.12, pp. 4655-4666, 2007.

[17] D. P. Wipf and B. D. Rao, "Sparse Bayesian learning for basis selection," *IEEE Transactions on Signal Processing*, vol.52, no.8, pp. 2153-2164, 2004.

[18] J. Gao, B. Deng, Y. Qin, H. Wang and X. Li, "Enhanced Radar Imaging Using a Complex-valued Convolutional Neural Network", arXiv, 2017.